\RequirePackage{fix-cm}
\documentclass{svjour3ujm} 
\smartqed  
\usepackage{graphicx}
\begin{document}

\title{Niels Bohr, objectivity, and the irreversibility of measurements}
\author{Ulrich J. Mohrhoff}
\institute{U. Mohrhoff\at
              Sri Aurobindo International Centre of Education \\
							Pondicherry 605002, India \\
              \\
							This is a post-peer-review, pre-copyedit version of an article published in Quantum Studies: Mathematics and Foundation. The final authenticated version is available online at https://doi.org/10.1007/s40509-019-00213-6.}

\maketitle
\begin{abstract}
The only acceptable reason why measurements are irreversible and outcomes definite is the intrinsic definiteness and irreversibility of human sensory experience. While QBists deserve credit for their spirited defense of this position, Niels Bohr urged it nearly a century ago, albeit in such elliptic ways that the core of his message has been lost or distorted beyond recognition. Then as now, the objectivity of empirical science was called into question. It was defended by Bohr along the lines of Kant's (then) revolutionary theory of science, according to which the possibility of empirical science hinges on the possibility of thinking of experiences as experiences of a system of interacting, re-identifiable objects. What Bohr added to Kant's theory was his insight that empirical knowledge was not necessarily limited to what is directly accessible to the senses, and that, therefore, it does not have to be solely a knowledge of objects of sensible intuition. It can also be a knowledge of phenomena that are not objects of sensible intuition but instead are constituted by experimental contexts, which are objects of sensible intuition. Bohr's grounding of objectivity (or the objectivity consistent with quantum mechanics), however, is weaker than Kant's (or the objectivity consistent with Newtonian physics). This conclusion is based on an examination of Bohr's intentions in appealing to irreversible amplification effects or the sufficient size and weight of the measurement apparatus.
\keywords{Bohr \and experience \and irreversibility \and Kant \and measurement \and objectivity \and QBism}
\end{abstract}
\section{Introduction}
\label{intro}
The beginning of the 21st Century saw the launch of a radically epistemic interpretation of quantum mechanics, by Carlton Caves, Chris Fuchs, and Ruediger Schack \cite{CFS2002}. Initially conceived as a generalized personalist Bayesian theory of probability called ``Quantum Bayesianism,'' it has since been re-branded as ``QBism,'' the term David Mermin \cite{MerminQBnotCop} prefers, considering it ``as big a break with 20th century ways of thinking about science as Cubism was with 19th century ways of thinking about art.'' The big break lies not in the emphasis that the mathematical apparatus of quantum mechanics is a probability calculus but in this \emph{plus} a radically subjective Bayesian interpretation of probability \emph{plus} a radically subjective interpretation of the events to which, and on the basis of which, probabilities are assigned by ``users'' (of quantum mechanics) or ``agents'' (in a quantum world).

Mermin lists four stages of acceptance a radical new idea goes through: (1)~It's nonsense; (2)~it's well known; (3)~it's trivial; (4)~I thought of it first. He holds that by now Stage~(2) is well underway. A growing minority of philosophically-minded physicists are beginning to maintain that there is nothing very new in QBism, that it is in fact just another version of the Copenhagen interpretation. I, too, started out at Stage~(1), but have now reached an unlisted stage. I believe that Niels Bohr came tantalizingly close to adopting a QBist stance, and that the principal differences between his views and QBism are attributable to the fact that he wrote before interpreting quantum mechanics became a growth industry, while QBism emerged in reaction to an ever-growing number of attempts at averting the ``disaster of objectification'' \cite{vF1990}. 

I am indebted to QBism for two reasons: (i)~it made me look more closely at Bohr's original writings (and discover that the various ``Copenhagen interpretations'' are more or less grotesque caricatures of his views), and (ii)~it made me come round to seeing that there is no difference between observations qua experiences and observations qua measurement outcomes (which solves the legitimate version of the measurement problem by explaining why measurements are irreversible and outcomes definite). 

The aim of the present paper is to defend the thesis that Bohr and QBism are on the same page when it comes to accounting for the irreversibility of measurements and the definiteness of outcomes: Bohr, too, attributed these incontrovertible features of human sensory experience to the intrinsic definiteness and irreversibility of human sensory experience. This raises the obvious question as to the objectivity of empirical science in general and of quantum mechanics in particular. The first philosopher to address this question (with regard to Newtonian mechanics) was Immanuel Kant, while the first physicist to address this question (with regard to quantum mechanics) was Niels Bohr.

While there are certain passages in Bohr's writings which make said thesis seem compelling, there are others---and sometimes the very same passages---that make it seem as if Bohr attributed the irreversibility of measurements to irreversible processes or such ``macroscopic'' features as the size and weight of the measurement apparatus. I shall examine these passages with a view to clarifying Bohr's intentions.

My first order of business, carried out in Sec.~\ref{QBism}, is to dispel a certain misconception about Bohr's intentions, which appears to be shared by advocates of QBism. Section~\ref{Kant} then outlines Kant's theory of science, which strongly influenced the prevalent philosophical attitude of physicists well into the first quarter of the 20th Century. Section~\ref{enterqm} sketches the apparent collapse of Kantism due to the advent of quantum mechanics, and it explains why Bohr seems obscure today. In Sec.~\ref{qupha} I~argue that Bohr basically accepted Kant's meaning of objectivity, which hinges on the possibility of thinking of experiences (qua phenomena or appearances) as experiences of objects, but extended it to include quantum phenomena: ``the objective character of the description in atomic physics depends on the detailed specification of the experimental conditions under which evidence is gained'' [BCW\,10:215] \cite{BCW}. Section~\ref{obir}, finally, addresses the issues at the heart of the present inquiry, as to how Bohr's intentions in appealing to irreversible amplification effects and sufficiently large and heavy measuring instruments ought to be understood.

More extensive discussions of the issues surrounding the particular focus of the present paper are available \cite{B4Bohr,BQBandB}.

\section{QBism: a minimal critique}
\label{QBism}
While QBism, as propounded by Fuchs, Mermin, and Schack \cite{FMS2014},  agrees with Bohr ``that the primitive concept of \emph{experience} is fundamental to an understanding of science,'' it partakes of current misconceptions regarding Bohr's mature views, which according to Clifford Hooker \cite{Hooker1972}  ``remained more or less stable at least over the latter thirty years of Bohr's life'' (i.e., since at least 1932). The following assertion by Mermin \cite{MerminQBnotCop} may serve as a representative example:
\begin{quote}
Niels Bohr \dots\ delivered some remarkably QBist-sounding pronouncements, though by ``experience'' I believe he meant the objective readings of large classical instruments and not the personal experience of a particular user of quantum mechanics.
\end{quote}
As there are more senses of  ``experience'' than the two adduced, what is asserted here amounts to a false dichotomy. A third possible sense is the shared experience of the scientific community, to which Mermin alluded a couple of paragraphs earlier, where he refers to ``the common external world we have all negotiated with each other'':
\begin{quote}
Those who reject QBism \dots\ reify the common external world we have all negotiated with each other, purging from the story any reference\break to the origins of our common world in the private experiences we try to share with each other through language. 
\end{quote}
Mermin thus is clearly aware of the difference between reification and objectivation. Whereas the reification of our common external world subsumes the belief that the world we perceive exists independently of our perceptions, that the world we describe exists independently of our descriptions, or that the world we mentally construct exists independently of our constructing minds, ``objectification'' is just another word for the collective negotiation of a common external world. Mermin approvingly quotes Schr\"odinger, who provided the following definition \cite{SchrLifeMindMatterPO}:
\begin{quote}
By [objectivation] I mean the thing that is also frequently called the ``hypothesis of the real world'' around us\dots.  Without being aware of it and without being rigorously systematic about it, we exclude the Subject of Cognizance from the domain of nature that we endeavour to understand. We step with our own person back into the part of an onlooker who does not belong to the world, which by this very procedure becomes an objective world.
\end{quote}
A little later in the same chapter, Schr\"odinger writes:
\begin{quote}
Mind has erected the objective outside world of the natural philosopher out of its own stuff. Mind could not cope with this gigantic task otherwise than by the simplifying device of excluding itself---withdrawing from its conceptual creation.
\end{quote}
This idea goes back to Kant, who was a major philosophical source for  Schr\"odinger.

``When speaking of a conceptual framework,'' Bohr wrote, ``we merely refer to an unambiguous logical representation of relations between experiences'' [BCW\break 10:84]. In Bohr's time and cultural environment, Kant's theory of science still exercised considerable influence. There can be little doubt that the conceptual framework Bohr had in mind was in all important respects the conceptual framework staked out by Kant. This provided the general structure of an object-oriented language, made up of logical representations of spatiotemporal relations between experiences.

Here is Bohr's ``remarkably QBist-sounding'' pronouncement quoted by Mermin right after expressing his belief that by ``experience'' Bohr meant the objective readings of large classical instruments and not the personal experience of a particular user of quantum mechanics: 
\begin{quote}
in our description of nature the purpose is not to disclose the real essence of the phenomena but only to track down, so far as it is possible, relations between the manifold aspects of our experience [BCW\,6:296].
\end{quote}
Most likely what Bohr meant by experience was (in the words of Carl Friedrich von Weizs\"acker \cite{vW2014}) that ``we can learn from the past for the future,'' that ``this learning can be formulated in terms of concepts,'' and that ``there are therefore things such as---to put it tentatively---recurring events which we can recognize with recognizable words.'' Von Weizs\"acker's definition highlights the inextricability of the subjective and objective aspects of experience. These recurring events occur in my subjective experience as well as in yours. The ability to recognize them with recognizable words, on the other hand, involves language. It implies that we can communicate them to each other, whereby they become objective, i.e., parts of our common external world. Combined with his insistence on ``the subjective character of all experience'' [BCW\,6:279; 7:259], Bohr's predication of the objectivity of the quantum theory upon language and communication (Sec.~\ref{obir} below) makes it rather immaterial whether by ``experience'' he meant the origin of our common external world in our subjective experiences or the common external world we have negotiated with each other through language. What he would not have denied for a moment is the origination of objective experience in our respective private experiences.

\section{An outline of Kant's theory of science}
\label{Kant}
Before Kant,  there appears to have been no philosopher who did not have a correspondence theory of truth. John Locke and Ren\'e Descartes distinguished between primary qualities, which existed ``out there'' and were properly represented in or by our minds, and secondary qualities, which bore no similarities to sensations (our modern ``qualia'') but nevertheless existed ``out there'' as something that had the power to produce sensations in the perceiving subject. George Berkeley then made it clear that to ask whether a table is the same size and shape as my mental image of it was to ask an absurd question. To Kant, therefore, \emph{all} qualities were secondary. Everything we say about an object is of the form: it has the power to affect us in such and such a way. Nothing of what we say about an object describes the object as it is in itself. Nor did Kant stop at saying that if I see a desk, there is a thing-in-itself that has the power to appear as a desk, and if I see a chair in front of the desk, there is another thing-in-itself that has the power to appear as a chair. For Kant, there was only one thing-in-itself, an empirically inaccessible reality that was \emph{not} the concern of empirical science. 

Relations between the manifold aspects of our ex\-perience---both spatiotemporal and logical or gramma\-tical---are the very means by which we, according to Kant, construct the objective world. The relation between a logical or grammatical subject and a predicate makes it possible for us to think of a particular nexus of perceptions as the properties of a \emph{substance}, connected to it as predicates are connected to a subject. It makes it possible for me to think of my perceptions as connected not by me, in my experience, but in an object ``out there.'' The logical relation between antecedent and consequent (if\dots\ then\dots) makes it possible for us to think of what we perceive at different times as properties of substances connected by \emph{causality}. It makes it possible for me to think of successive perceptions as connected not merely in my experience but also objectively, by a causal nexus ``out there.'' And the category of community or reciprocity, which Kant associated with the disjunctive relation (either\dots\ or\dots), makes it possible for us to think of what we perceive in different locations as properties of substances connected by a \emph{reciprocal action}. It makes it possible for me to think of simultaneous perceptions as connected not only in my experience but objectively.

But if we are to be able to think of perceptions as properties of substances, or as causally connected, or as affecting each other, the connections must be regular. For perceptions to be perceptions of a particular kind of thing (say, an elephant), they must be connected in an orderly way, according to a concept denoting a lawful concurrence of perceptions. For perceptions to be causally connected, like (say) lightning and thunder, they must fall under a causal law, according to which one perception necessitates the subsequent occurrence of another. (By establishing a causal relation falling under a causal law, we also establish an objective temporal relation.) And for perceptions to be reciprocally connected, like (say) the Earth and the Moon, they must affect each other according to a reciprocal law, such as Newton's law of gravity. It is through lawful connections in the ``manifold of appearances'' that we are able to think of appearances as perceptions of a self-existent system of causally evolving (and thus re-identifiable) objects, from which we, the experiencing subjects, can detach ourselves. 

By placing the subject of empirical science squarely into the context of human experience, Kant dispelled many qualms that had been shared by thinkers at the end of the 18th century---qualms about the objective nature of geometry, the purely mathematical nature of Newton's theory, the unintelligibility of action at a distance, Galileo's principle of relativity, to name a few. 

Around the beginning of the 20th Century, scientists and philosophers alike had realized that renouncing ontological prejudices and sticking to operationally definable notions was the safest way to arrive at reliable knowledge. At the same time classical physics, still deemed eminently successful, appeared to support a realistic interpretation. What made it possible to reconcile these opposing tendencies was Kant's theory of science. It offered a way to go on talking in realist language about, e.g., electromagnetic waves propagating in vacuum, while disavowing ontological inclinations. Kant's theory of science was therefore widely considered to be tightly linked with classical physics, and to make the latter philosophically acceptable \cite{dEspagnat2010}.

\section{Enter quantum mechanics}
\label{enterqm}
When quantum mechanics came along, this world view collapsed. For the present-day physicist, it is not easy to understand the bewilderment that the founders and their contemporaries experienced in the early days of quantum mechanics. In this new field of experience, the possibility of organizing sense impressions into interacting, re-identifiable objects (which for Kant was a precondition of the possibility of empirical  science) no longer seemed to exist. ``Atoms---our modern atoms, the ultimate particles---must no longer be regarded as identifiable individuals,'' Schr\"odinger wrote \cite{SchrNGSH}, ``This is a stronger deviation from the original idea of an atom than anybody had ever contemplated. We must be prepared for anything.'' Here is how Mermin's alter ego Mozart \cite{Mermin_epochs} remembered it: 
\begin{quote}
All the verities of the preceding two centuries, held by physicists and ordinary people alike, simply fell apart---collapsed. We had to start all over again, and we came up with something that worked just beautifully but was so strange that nobody had any idea what it meant except Bohr, and practically nobody could understand him. So naturally we kept probing further, getting to smaller and smaller length scales, waiting for the next revolution to shed some light on the meaning of the old one.
\end{quote}
That revolution never came. Quantum mechanics works as beautifully in the nucleus as it does in the atom; and it works as beautifully in the nucleon as it does in the nucleus, seven or eight orders of magnitude below the level for which it was designed. It also works beautifully many orders of magnitude above that level, as for example in a superconductor.

It is in fact not quite correct to say that practically nobody could understand Bohr. Historically, Bohr's reply \cite{BohrEPRreply} to the incompleteness argument of Einstein, Podolsky, and Rosen was taken as a definitive refutation by the physics community (although it may be said that no two members of the physics community understood him in quite the same way). It was only in the mid 1950s, when interpreting quantum mechanics came to mean grafting a metaphysical narrative onto a mathematical formalism in a language that is sufficiently vague philosophically to be understood by all and sundry, that Bohr's perspective was lost. His paper, which only treated the mathematical formalism in a footnote, was then seen as missing the point. 

There are three reasons (listed by Catherine Chevalley \cite{Chevalley99}) why Bohr seems obscure today. The first is that Bohr's views have come to be equated with one variant or another of the Copenhagen interpretation. The latter only emerged in the mid-1950's, in response to David Bohm's hidden-variables theory and the Marxist critique of Bohr's alleged idealism, which had inspired Bohm. The second reason is that Bohr's readers will usually not find in his writings what they expected to find, while they will find a number of things that they did not expect. What they expect is a take on problems arising from the reification of a probability calculus, such as the problem of objectification or the quantum-to-classical transition. What they find instead is discussions of philosophical issues such as the meanings of ``objectivity,''  ``truth,'' and ``reality'' and the roles played by language and communication. The third reason is the aforementioned ``dumbing down'' of what it means to make physical sense of quantum theory's mathematical formalism. For Bohr (as also for Werner Heisenberg and Wolfgang Pauli) it meant finding a conceptual framework that could take the place of the framework originally staked out by Kant.

\section{Quantum phenomena}
\label{qupha}
In the analytic part of his Critique \cite{KantCPR}, Kant argued that ``space and time are only forms of sensible intuition, and therefore only conditions of the existence of the things as appearances.'' (``Intuition'' is the standard translation of \emph{Anschauung}, which covers both visual perception and visual imagination.) In 1922 Bohr surmised as much in a letter to his philosophical mentor Harald H{\o}ffding: ``these difficulties are of such a kind that they hardly allow us to hope, within the world of atoms, to implement a description in space and time of the kind corresponding to our usual sensory images'' [BCW10:\,513]. By 1929 Bohr was convinced that  ``the facts which are revealed to us by the quantum theory \dots\ lie outside the domain of our ordinary forms of perception'' [BCW\,6:217], space and time. 

Kant went on to argue that
\begin{quote}
we have no concepts of the understanding and hence no elements for the cognition of things except insofar as an intuition can be given corresponding to these concepts, consequently that we can have cognition of no object as a thing in itself, but only insofar as it is an object of sensible intuition, i.e. as an appearance.
\end{quote}
Bohr could not have agreed more, insisting as he did that meaningful physical concepts have not only mathematical but also visualizable content. Such concepts are associated with pictures, like the picture of a particle following a trajectory or the picture of a wave propagating in space. In the classical theory, a single picture could accommodate all of the properties a system can have. When quantum theory came along, that all-encompassing picture fell apart. Unless certain experimental conditions obtained, it was impossible to picture the electron as following a trajectory (which was nevertheless a routine presupposition in setting up Stern--Gerlach experiments and in interpreting cloud-chamber photographs), and there was no way in which to apply the concept of position. And unless certain other, incompatible, experimental conditions obtained, it was impossible to picture the electron as a traveling wave (which was nevertheless a routine presupposition in interpreting the scattering of electrons by crystals), and there was no way in which to apply the concept of momentum.

Bohr settled on the nexus between pictures, physical concepts, and experimental arrangements as key to ``the task of bringing order into an entirely new field of experience'' \cite{BohrSchilpp}. If the visualizable content of physical concepts cannot be described in terms of compatible pictures, it has to be described in terms of something that \emph{can} be so described, and what can be so described are the experimental conditions under which the incompatible physical concepts can be employed. What distinguishes these experimental conditions from the quantum systems under investigation is their accessibility to direct sensory experience.

Bohr basically accepted Kant's view that objectivity hinges on the possibility of thinking of appearances as experiences of interacting, re-identifiable objects. What Bohr added to Kant's theory of science was his insight that empirical knowledge was not necessarily limited to what is \emph{directly} accessible to our senses, and that, therefore, it does not have to be \emph{solely} a knowledge of objects of sensible intuition. It can also be a knowledge of things that are \emph{not} objects of sensible intuition but instead are constituted by experimental conditions, which themselves \emph{are} objects of sensible intuition: ``the objective character of the description in atomic physics depends on the detailed specification of the experimental conditions under which evidence is gained'' [BCW\,10:215]. Quantum mechanics does not do away with objects of sensible intuition but supplements them with \emph{quantum phenomena}:
\begin{quote}
all unambiguous interpretation of the quantum mechanical formalism involves the fixation of the external conditions, defining the initial state of the atomic system concerned and the character of the possible predictions as regards subsequent observable properties of that system. Any measurement in quantum theory can in fact only refer either to a fixation of the initial state or to the test of such predictions, and it is first the combination of measurements of both kinds which constitutes a well-defined phenomenon. [BCW 7:312]
\end{quote}

\section{Objectivity and the issue of irreversibility}
\label{obir}
If the terminology of quantum phenomena is used consistently, then nothing---at any rate, nothing we know how to think about---happens between ``the fixation of the external conditions, defining the initial state of the atomic system concerned'' and ``the subsequent observable properties of that system.'' 
Any story purporting to detail a course of events in the interval between a system preparation and a subsequent observation is inconsistent with ``the essential wholeness of a quantum phenomenon,'' which ``finds its logical expression in the circumstance that any attempt at its subdivision would demand a change in the experimental arrangement incompatible with its appearance'' [BCW\,10:278]. What, then, are we to make of the following passages [emphases added]?
\begin{quote}
Every well-defined atomic phenomenon is closed in itself, since its observation implies a permanent mark on a photographic plate \emph{left by the impact of an electron} or similar recordings \emph{obtained by suitable amplification devices of essentially irreversible functioning}. [BCW\,10:89]

\smallskip Information concerning atomic objects consists solely in the marks they make on these measuring instruments, as, for instance, a spot \emph{produced by the impact of an electron on a photographic plate} placed in the experimental arrangement. The circumstance that such marks are \emph{due to irreversible amplification effects} endows the phenomena with a peculiarly closed character pointing directly to the irreversibility in principle of the very notion of observation. [BCW\,10:120]

\smallskip In this connection, it is also essential to remember that all unambiguous information concerning atomic objects is derived from the permanent marks---such as a spot on a photographic plate, \emph{caused by the impact of an electron}---left on the bodies which define the experimental conditions. Far from involving any special intricacy, the\emph{ irreversible amplification effects on which the recording of the presence of atomic objects rests} rather remind us of the essential irreversibility inherent in the very concept of observation. [BCW\,7:390]
\end{quote}
If a well-defined atomic phenomenon is closed, how can something happen between the fixation of the external conditions and a permanent mark on a photographic plate? Does not the interposition of the impact of an electron and subsequent amplification effects amount to a subdivision of the phenomenon in question? 

Ole Ulfbeck and Aage Bohr \cite{UlfbeckBohr} have shed light on this issue. For them,  clicks in counters are ``events in spacetime, belonging to the world of experience.'' While clicks can be classified as electron clicks, neutron clicks, etc., ``there are no electrons and neutrons on the spacetime scene'' and
\begin{quote}
there is no wave function for an electron or a neutron but [only] a wave function for electron clicks and neutron clicks. \dots there is no longer a particle passing through the apparatus and producing the click. Instead, the connection between source and counter is inherently non-local.
\end{quote}
 The key to resolving the issue at hand is that each click has an onset---``a beginning from which the click evolves as a signal in the counter.'' This onset 
\begin{quote}
has no precursor in spacetime and, hence, does not belong to a chain of causal events. In other words, the onset of the click is not the effect of something, and it has no meaning to ask how the onset occurred.

\smallskip [T]he occurrence of genuinely fortuitous clicks, coming by themselves, is recognized as the basic material that quantum mechanics deals with. 

\smallskip [T]he wave function enters the theory not as an independent element, but in the role of encoding the probability distributions for the clicks. 

\smallskip [T]he steps in the development of the click, envisaged in the usual picture, are not events that have taken place on the spacetime scene.

\smallskip [T]he downward path from macroscopic events in spacetime, which in standard quantum mechanics continues into the regime of the particles, does not extend beyond the onsets of the clicks.
\end{quote}
If irreversible amplification effects---the steps in the development of the click---only occur ``in the usual picture,'' then they neither modify nor subdivide the quantum phenomenon in which---through an illegitimate extension of the object-oriented language of classical physics---they are said to occur.

For Niels Bohr, ``the physical content of quantum mechanics [was] exhausted by its power to formulate statistical laws governing observations obtained under conditions specified in plain language'' [BCW\,10:159]. A quantum phenomenon thus has a statistical component, which correlates events in the world of experience. The so-called irreversible amplification effects belong to this statistical component. The unmediated step from the source to the onset of the click, and the subsequent unmediated steps in the development of the click, are steps in a gazillion of alternative sequences of possible outcomes of \emph{unperformed} measurements, and unperformed measurements do not affect the essential wholeness of a quantum phenomenon.

Niels Bohr, moreover, strongly cautioned against the terminology of ``disturbing a phenomenon by observation'' and of ``creating physical attributes to objects by measuring processes'' [BCW\,7:316]. If there is nothing to be disturbed by observation, if even the dichotomy of objects and attributes created for them by measuring processes is unwarranted, then it is not just the measured property that is constituted by the experimental conditions under which it is observed; it is the quantum system itself that is so constituted. Recently this point was forcefully made by Brigitte Falkenburg in her monograph \emph{Particle Metaphysics} \cite{Falkenburg2007-205f}:
\begin{quote}
[O]nly the experimental context (and our ways of conceiving of it in classical terms) makes it possible to talk in a sloppy way of \emph{quantum objects}\dots. Bare quantum ``objects'' are just bundles of properties which underlie superselection rules and which exhibit non-local, acausal correlations\dots. They seem to be Lockean empirical substances, that is, collections of empirical properties which constantly go together. However, they are only individuated by the experimental apparatus in which they are measured or the concrete quantum phenomenon to which they belong\dots. They can only be individuated as context-dependent quantum \emph{phenomena}.\break Without a given experimental context, the reference of quantum concepts goes astray. In this point, Bohr is absolutely right up to the present day.
\end{quote}

In the following passages [emphases added], Bohr goes beyond invoking irreversible amplification effects, apparently arguing that the quantum features involved in the atomic constitution of a measurement apparatus (or the statistical element in its description) can be neglected because the relevant parts of a measurement apparatus are sufficiently large and heavy.
\begin{quote}
In actual experimentation this demand [that the experimental arrangement as well as the recording of observations be expressed in the common language] is met by the specification of the experimental conditions by means of bodies like diaphragms and photographic plates \emph{so large and heavy that the statistical element in their description can be neglected}. The observations consist in the recording of permanent marks on these instruments, and the fact that the amplification devices used in the production of such marks involves essentially irreversible processes presents no new observational problem, but merely stresses the element of irreversibility inherent in the definition of the very concept of observation. [BCW\,10:212]

\smallskip In actual physical experimentation this requirement [that we must employ common language to communicate what we have done and what we have learned by putting questions to nature in the form of experiments] is fulfilled by using as measuring instruments rigid bodies like diaphragms, lenses, and photographic plates \emph{sufficiently large and heavy to allow an account of their shape and relative positions and displacements without regard to any quantum features inherently involved in their atomic constitution}. [BCW\,10:165]
\end{quote}
How can the size and weight of a measuring device justify
\begin{itemize}
\item[---] the irreversibility in principle of the very notion of observation [BCW\,10:120],
\item[---] the essential irreversibility inherent in the very concept of observation [BCW 7:390; BCW\,10:128],
\item[---] the irreversibility which is implied in principle in the very concept of observation [BCW\,10:165], or
\item[---] the element of irreversibility inherent in the definition of the very concept of observation [BCW 10:212]?
\end{itemize}
As my engagement with QBism made me realize that only the incontestable irreversibility of human sensory experience can justify the irreversibility of observations, so my subsequent engagement with the Bohr canon convinced me that this is what Bohr himself had been trying to say. For Bohr, ``the emphasis on the subjective character of the idea of observation [was] essential'' [BCW\,10:496]. If the description of atomic phenomena nevertheless has ``a perfectly objective character,'' it was ``in the sense that no explicit reference is made to any \emph{individual} observer and that therefore \dots\ no ambiguity is involved in the communication of information'' [BCW\,10:128, emphasis added]. It was never in the sense that no reference was made to the community of communicating observers or to the incontestable irreversibility of their experiences.

Like Kant, Bohr was concerned with the possibility of an objective knowledge of phenomena (or appearances, or experiences). Kant primarily addressed the possibility of organizing appearances into a world of objects. This involved concepts that allowed the individual not only to think of appearances as a world of external objects but also to communicate with others about a world of external objects. Bohr primarily addressed the requisite possibility of communication. This involved the use of concepts that allowed the individual not only to communicate with others about a world of external objects but also to think of appearances as a world of external objects. Bohr and Kant were on the same page.

Bohr's argument, if it can be put in a nutshell, was
\begin{quote}
\dots that by the word ``experiment'' we refer to a situation where we can tell others what we have done and what we have learned and that, therefore, the account of the experimental arrangement and of the results of the observations must be expressed in unambiguous language with suitable application of the terminology of classical physics. [BCW\,7:349]
\end{quote}
Two expressions are significant here: ``unambiguous language'' and ``the terminology of classical physics.'' At present (August 2019) a combined Google search for ``Bohr'' and ``classical language'' (the latter term including the quotes) yields nearly 5,000 results, while a Google search for ``Bohr'' and ``language of classical physics'' yields more than 24,000 results. By contrast, searching the 13 volumes of the \emph{Complete Works} of Niels Bohr does not yield a \emph{single} occurrence of either ``classical language'' or ``language of classical physics.'' It is the ubiquity in the secondary literature of these latter expressions that is chiefly responsible for the wide\-spread misconceptions about Bohr's thinking.

While Bohr insisted on the use of classical \emph{concepts} (or the terminology of classical physics, or simply ``elementary physical concepts'' [BCW\,7:394]), the \emph{language} on the use of which he insisted was ``ordinary language'' [BCW\,7:355], ``plain language'' [BCW\,10:159], the ``common human language'' [BCW\,10:157--158], or the ``language common to all'' [BCW\,10:xxxvii]. A distinction must therefore be drawn between the role that classical concepts played in Bohr's thinking and the role that was played by the common human language. The common human language is the object-oriented language of everyday discourse, while classical concepts are not proprietary to classical physics but denote attributes that owe their meanings to our forms of perception, such as position, orientation, and the ones that are defined in terms of invariances under spacetime transformations.

One day during tea at his institute, Bohr was sitting next to Edward Teller and Carl Friedrich von Weiz\-s\"acker. Von Weizs\"acker \cite{vW_Structure} recalls that when Teller suggested that ``after a longer period of getting accustomed to quantum theory we might be able after all to replace the classical concepts by quantum theoretical ones,'' Bohr listened, apparently absent-mindedly,  and said at last: ``Oh, I understand. We also might as well say that we are not sitting here and drinking tea but that all this is merely a dream.'' If we are dreaming, we are unable to tell others what we have done and what we have learned. Hence, ``it would be a misconception to believe that the difficulties of the atomic theory may be evaded by eventually replacing the concepts of classical physics by new conceptual forms'' [BCW\,6:294].

What distinguishes the objectivity that could be achieved in the age of Kant from the objectivity that can be achieved in the age of quantum mechanics is that a complete elision of the subject is no longer feasible. Having asserted that ``we can have cognition of no object as a thing in itself, but only insofar as it is an object of sensible intuition, i.e. as an appearance,'' Kant went on to affirm that
\begin{quote}
even if we cannot \emph{cognize} these same objects as things in themselves, we at least must be able to \emph{think} them as things in themselves. For otherwise there would follow the absurd proposition that there is an appearance without anything that appears. \cite{KantCPR}
\end{quote}
If the only relevant context is human experience, or if the reach of human sensory experience is unlimited (as classical physics takes it to be), the elision of the subject can be achieved. It is possible to transmogrify calculational tools into objective physical processes with some measure of consistency \cite{Mermin_badhabit}. But if the experimental context is relevant as well, or if the reach of human sensory experience is limited, then the elision of the subject is a lost cause, and so is the transmogrification of calculational tools into physical mechanism or natural processes. If one nevertheless wants to establish the irreversibility of measurements and the definiteness of outcomes without invoking the definiteness and irreversibility of human sensory experience, one has no choice but to invoke such ``macroscopic'' features as the size and weight of a measurement apparatus. This (to appropriate a well-known passage by John Bell) ``is like a snake trying to swallow itself by the tail. It can be done---up to a point. But it becomes embarrassing for the spectators even before it becomes uncomfortable for the snake'' \cite{Bell90}.

In the words of Bernard d'Espagnat \cite{dEspagnat2010}, quantum mechanics practically compels us
\begin{quote}
to adopt the idea that was, in fact, at the very core of Kantism and constitutes its truly original contribution to philosophical thinking, to wit, the view that things and events, far from being elements of a ``reality per se,'' are just phenomena, that is, elements of our experience.
\end{quote}
There remains the issue of Bohr's intentions in invoking, in lieu of the irreversibility of human experience, the size and weight of the measurement apparatus. Did he do it to appease the na\"{\i}ve realistic inclinations of lesser minds? Or did he do it in order to indicate the price one had to pay (but which he personally was not willing to pay) for exorcising every possible reference to human sensory experience? My suggested answer is ``yes'' to both questions.


\begin{thebibliography}{99}
\bibitem{CFS2002} 
Caves, C.M., Fuchs, C.A., Schack, R.: Quantum probabilities as Bayesian probabilities.
Phys. Rev. A 65, 022305 (2002)

\bibitem{MerminQBnotCop} 
Mermin, N.D.: Why QBism is not the Copenhagen interpretation and what John Bell might have thought of it. In R. Bertlmann and A. Zeilinger (eds.), Quantum [Un]Speakables II: 50 Years of Bell's Theorem, pp. 83--93. Springer, Berlin (2017)

\bibitem{vF1990} 
Van Fraassen, B.C.: The problem of measurement in quantum mechanics. In P. Lahti, P. Mittelstaedt (eds.), Symposium on the Foundations of Modern Physics, pp. 497--503. World Scientific, Singapore (1990).

\bibitem{BCW} 
[BCW\,x:y] refers to volume and page number(s) of Rosenfeld, L. (general editor), Niels Bohr: Collected Works in 13 Volumes. Elsevier, Amsterdam, Netherlands (1972--2008)

\bibitem{B4Bohr}
Mohrhoff, U.: ``B'' is for Bohr, to appear in the combined Proceedings of the Workshops on Meaning and Structure of Quantum Mechanics at Buenos Aires (2016 and 2019), arXiv:1905.07118 [quant-ph]

\bibitem{BQBandB}
Mohrhoff, U.: Bohr, QBism, and beyond, to appear in the Proceedings of the 2019 Copenhagen Interpretation and Beyond Conference at Chapman University, Los Angeles, honoring Henry P. Stapp, arXiv:1907.11405 [quant-ph]

\bibitem{FMS2014} 
Fuchs, C.A., Mermin, N.D., Schack, R.: An introduction to QBism with an application to the locality of quantum mechanics.  Am. J. Phys. 82(8), 749--754 (2014).

\bibitem{Hooker1972} 
Hooker, C.A.: The nature of quantum mechanical reality: Einstein versus Bohr. In R.G. Colodny (ed.), Paradigms \&\ Paradoxes: The Philosophical Challenge of the Quantum Domain, pp. 67--302. University of Pittsburgh Press, Pittsburgh (1972)

\bibitem{SchrLifeMindMatterPO} 
Schr\"odinger, E.: The Principle of Objectivation. In What Is Life? With: Mind and Matter \&\ Autobiographical Sketches, pp. 117--127. Cambridge University Press, Cambridge, U.K. (1992)

\bibitem{vW2014}
Von Weizs\"acker, C.F.: The Unity of Physics as a Philosophical Problem. In M. Drieschner (ed.), Carl Friedrich von Weizs\"acker: Major Texts in Physics, pp. 67--71. Springer, Heidelberg (2014)

\bibitem{dEspagnat2010}
D'Espagnat, B.: A Physicist's Approach to Kant. In M. Bitbol, P. Kerszberg, J. Petitot (eds.), Constituting Objectivity: Transcendental Perspectives on Modern Physics, pp. 481--490. Springer, Berlin (2009)

\bibitem{SchrNGSH} 
Schr\"odinger, E.: Nature and the Greeks, and Science and Humanism, p.~162. Canto Classics, Cambridge, UK (2014)

\bibitem{Mermin_epochs} 
Mermin, N.D.: What's wrong with those epochs. Phys. Today 43(11), 9--11 (1990)

\bibitem{BohrEPRreply} 
Bohr, N.: Can quantum-mechanical description of physical reality be considered complete? Phys. Rev. 48, 696--702 (1935)

\bibitem{Chevalley99} 
C. Chevalley, C.: Why do we find Bohr obscure? In D. Greenberger, W.L. Reiter, A. Zeilinger (eds.), Epistemological and Experimental Perspectives on Quantum Physics, pp. 59--73. Kluwer, Dordrecht, Netherlands (1999)

\bibitem{KantCPR} 
Kant, I.: Critique of Pure Reason, transl. and ed. P. Guyer, A.W. Wood, pp. 115. Cambridge University Press, Cambridge, U.K. (1998)

\bibitem{BohrSchilpp} 
Bohr, N.: Discussion with Einstein on the epistemological problems in atomic physics. In P.A. Schilpp (ed.), Albert Einstein: Philosopher--Scientist, pp. 201--241. MFJ Books, New York (1949)

\bibitem{UlfbeckBohr} 
Ulfbeck, O., Bohr, A.: Genuine Fortuitousness. Where Did That Click Come From? Found. Phys. 31(5), 757--774 (2001)

\bibitem{Mermin_badhabit}
Mermin, N.D.: What's bad about this habit. Phys. Today 62(5), 8--9 (2009)

\bibitem{Bell90} 
Bell, J.S.: Against ``measurement.'' Phys. World , 33--40 (August 1990)

\bibitem{Falkenburg2007-205f} 
Falkenburg, B.: Particle Metaphysics: A Critical Account of Subatomic Reality. Springer (2007), pp.~205--206.

\bibitem{vW_Structure} 
Von Weizs\"acker, C.F.: The Structure of Physics, p. 25. Springer, Berlin (2006) 
\end{thebibliography}
\end{document}